# The positive piezoconductive effect in graphene

Kang Xu[1]*, Ke Wang[2,3]*, Wei Zhao[1]*, Wenzhong Bao[4,5], Erfu Liu[1], Yafei Ren[2,3], Miao Wang[1], Yajun Fu[1], Junwen Zeng[1], Zhaoguo Li[1], Wei Zhou[1], Fengqi Song[1], Xinran Wang[6], Yi Shi[6], Xiangang Wan[1], Michael S. Fuhrer[4,7], Baigeng Wang[1], Zhenhua Qiao[2,3], Feng Miao[1] & Dingyu Xing[1]

**As the thinnest conductive and elastic material, graphene is expected to play a crucial role in post-Moore era. Besides applications on electronic devices, graphene has shown great potential for nano-electromechanical systems. While interlayer interactions play a key role in modifying the electronic structures of layered materials, no attention has been given to their impact on electromechanical properties. Here we report the positive piezoconductive effect observed in suspended bi- and multi-layer graphene. The effect is highly layer number dependent and shows the most pronounced response for tri-layer graphene. The effect, and its dependence on the layer number, can be understood as resulting from the strain-induced competition between interlayer coupling and intralayer transport, as confirmed by the numerical calculations based on the non-equilibrium Green's function method. Our results enrich the understanding of graphene and point to layer number as a powerful tool for tuning the electromechanical properties of graphene for future applications.**

[1] National Laboratory of Solid State Microstructures, School of Physics, Collaborative Innovation Center of Advanced Microstructures, Nanjing University, Nanjing 210093, China.

[2] ICQD, Hefei National Laboratory for Physical Sciences at Microscale, and Synergetic Innovation Center of Quantum Information and Quantum Physics, University of Science and Technology of China, Hefei, Anhui 230026, China.

[3] Key Laboratory of Strongly-Coupled Quantum Matter Physics, Chinese Academy of Sciences, and Department of Physics, University of Science and Technology of China, Hefei, Anhui 230026, China.

[4] Department of Physics, University of Maryland, College Park, Maryland 20742,

USA.

[5] State Key Laboratory of ASIC and System, Department of Microelectronics, Fudan University, Shanghai 200433, China.

[6] School of Electronic Science and Engineering, Nanjing University, Nanjing 210093, China.

[7] School of Physics, Monash University, Monash, Victoria 3800, Australia.

\* These authors contributed equally to this work.

Correspondence and requests for materials should be addressed to F. M. (email: miao@nju.edu.cn), Z. Q. (email: qiao@ustc.edu.cn) or to B. W. (email: bgwang@nju.edu.cn).

Graphene[1-4] is an ideal material candidate for nano-electromechanical systems (NEMS)[5] due to many advantageous features, including unparalleled breaking length[6], ultrahigh carrier mobility[7], and excellent controllability of electronic structures via mechanical strain[8,9]. Many intriguing phenomena have been experimentally observed on strained graphene[10-12], including the observation of pseudo-magnetic fields exceeding 300 Tesla[13]. More fascinating phenomena have been theoretically predicted for strained graphene, but yet to be realized experimentally, such as the zero-field quantum Hall effect[14], strain induced superconductivity of graphene[15], and potential applications in valleytronics[16,17]. Thus, strain-controllable transport measurements are critical for in-depth understanding and further applications of graphene.

*In-situ* piezoconductive measurements on graphene provide an effective approach to study the correlations between electrical properties and mechanical strains. So far, studies have been focused on mono-layer graphene, and the negative piezoconductive effect has been widely reported, independent of the differences in graphene film synthesis, transfer methods, and sample substrates[18-23]. Systematic studies of the electromechanical properties of graphene with different number of layers and alterable interlayer interactions have been lacking.

Here we investigate the piezoconductive effect of suspended graphene membranes with various layer numbers by applying *in-situ* stress with a scanning probe. We observe positive piezoconductance in bi- and multi-layer graphene, with tri-layer graphene showing the most pronounced response. This intriguing phenomenon can be explained by the model of strain-induced competition between electronic interlayer coupling and intralayer transport, and further confirmed by numerical calculations based on non-equilibrium Green's function method.

**Results**

**Device fabrication and piezoconductive measurements.** Suspended graphene membranes are mechanically exfoliated and deposited on Si/SiO$_2$ wafers with pre-etched trenches. The number of the graphene layers is identified via color interference, and confirmed by Raman spectroscopy. Metal electrodes (5nm Ti/50nm Au) are deposited through home-made shadow masks, whi ch effectively avoid wet process-induced device performance degradation[24]. A typical device image is shown in Fig. 1a. To perform *in-situ* piezoconductive measurements, we introduce pressure-modulated conductance microscopy (PCM)[25,26], which utilizes a non-conducting atomic force microscopy (AFM) tip to apply adjustable local pressure on the top of the suspended graphene, yielding a topography image of the strained graphene. The conductance/resistance of the device is monitored simultaneously, and comparison of the conductance/resistance image and topography offers piezoconductive information of the suspended graphene membranes. The detailed experimental setup is schematically shown in Fig. 1b.

The measurements are carried out on mono- and multi-layer suspended graphene devices. Typical piezoconductive results from mono-, bi-, tri-, and tetra-layer devices are shown in Fig. 2a, 2b, 2c, and 2d respectively. For each figure, the left panel shows a line trace of the topography image, and the right panel shows the corresponding relative conductance change represented by $g$. Here $g(x)$ is defined by $\frac{G(x)-G_0}{G_0}$, where $G(x)$ is the device conductance at AFM tip position $x$ (where local pressure is applied), and $G_0$ is the undisturbed conductance value. For mono-layer suspended graphene, the device conductance drops upon local pressure applied (Fig. 2a), indicating negative piezoconductive (i.e. positive piezoresistive) effect. This is consistent with all previous studies[19,20]. However, for multi-layer suspended graphene, device conductance jumps upon local pressure applied (Fig. 2b-2d), indicating positive piezoconductive effect, which has never been reported on graphene devices. We note that the measured tri-layer graphene device is stacked with the common Bernal (ABA) structure, as confirmed by Raman spectroscopy.

To further study the observed layer number dependent piezoconductive effect, we perform the same measurement on various suspended graphene devices (layer number $n = 1, 2, 3, 4, 6$) with different strains (up to 1‰). The detailed results are shown in Fig. 2e. Here we plot the maximum relative conductance change $g_{\max}$ (which usually appears when AFM tip approaches the center of the suspended membranes). According to the geometry of our devices, the strain $\varepsilon$ can be calculated by the equation

$$\varepsilon = 2\frac{h^2}{l^2} \quad (1)$$

where $h$ is the maximum strain-induced deflection, and $l$ is the length of the suspended graphene (same as the width of the trench, around 3μm). Here $h$ has been corrected by subtracting the height of the tip at which the force begins to rise (which can be extracted from the deflection-force curve, see Supplementary Fig. 1 and Supplementary Note 1 for details). As shown in the $g_{\max}$ vs. $\varepsilon$ plot in Fig. 2e, several data points from a mono-layer graphene device fall in the negative regime. The gauge factor (usually defined by relative resistance change divided by strain) is estimated to be about 0.6, similar to the values previously reported[19,20]. In sharp contrast, all multi-layer graphene devices show positive conductance response. More interestingly, for similar strain value, the $n = 3$ (tri-layer) device shows much larger $g_{\max}(\varepsilon)$ than the other multi-layer devices ($n = 2, 4, 6$).

**Theoretical interpretation of the underlying physical origin**

The positive piezoconductive effect is difficult to explain in terms of strain induced decrease of Fermi velocity, the model which has been applied to strained mono-layer graphene[19,20]. In the system of back-gated suspended mono-layer graphene, a model of inhomogeneous carrier density redistribution predicted a positive piezoconductive effect[27] in contrast to our observation of negative piezoconductance for mono-layer graphene, and cannot explain our observation as well. The fact that positive piezoconductance is only observed in bi- and multi-layer graphene suggests that interlayer coupling is key; indeed, strain should modify the interaction between graphene layers, modifying the band structure and hence transport properties.

In order to understand the observed positive piezoconductive effect, we numerically calculate the transport properties of strained multi-layer graphene devices, which can be calculated by applying non-equilibrium Green's function technique and using two-terminal Landauer-Büttiker formula, with details described in Methods section. To numerically study the piezoconductive effect in the multi-layer graphene systems in the presence of an external pressure, we consider a $\pi$-orbital tight-binding model Hamiltonian, which is written as[4,28,29]:

$$H = H_0 + H_D + H_S \quad (2)$$

$$H_0 = -\mu \sum_i c_i^\dagger c_i + \sum_{\langle ij \rangle^{\text{intra,inter}}} \left(c_i^\dagger t_{\text{intra,inter}} c_j + H.c.\right) \quad (3)$$

$$H_D = \sum_i c_i^\dagger \varepsilon_i c_i + \sum_{ik} \left[a_{ik}^\dagger \varepsilon_k a_{ik} + \left(a_{ik}^\dagger t_k c_i + H.c.\right)\right] \quad (4)$$

$$H_S = \sum_i c_i^\dagger u_i c_i + \sum_{\langle ij \rangle^{\text{intra,inter}}} \left(c_i^\dagger \delta t_{ij}^{\text{intra,inter}} c_j + H.c.\right) \quad (5)$$

Where $c_i^\dagger$ and $c_i$ are the creation and annihilation operators on the site $i$. The first term $H_0$ describes the pristine multi-layer graphene sheets, with $\mu$ being the chemical potential and $t_{\text{intra,inter}}$ denoting the intra- and interlayer nearest neighbor hopping strength respectively. In our consideration, we take $t_{\text{intra}} = 2.60\text{eV}$ and $t_{\text{inter}} = 0.34\text{eV}$, respectively. The second term $H_\text{D}$ describes the influences of external disorders and the dephasing effect that is used to recover the macroscopic behavior from a finite-sized quantum system and can be modeled by attaching individual virtual leads at each site $i$. Here $\varepsilon_i$ is the on-site Anderson disorder strength that is uniformly distributed in the interval of [-*W*/2, *W*/2] with *W* characterizing the strength of disorder, and $a_{ik}^\dagger$ and $a_{ik}$ are the creation and annihilation operators for the virtual lead attached at the *i*-th site. In our consideration, we set the disorder strength to be 0.1eV, because the prepared multi-layer graphene sheets are rather clean and it can better smear off the room temperature induced thermal fluctuation. The last term $H_\text{S}$ represents the applied strain-induced effects including the induced site potentials and the variation of the intra- and interlayer nearest-neighbor hopping energies in the lattice-deformed region. The AFM tip-contacted lattices should experience noticeable lattice deformations, while the lattice deformation farther away from the AFM tip gradually decreases. Thus it is reasonable to assume that the variation of the intra- and interlayer hopping energy due to the lattice deformation takes the form of $\delta t_{ij}^{\text{intra,inter}} = \delta t_0 \left(1 - \frac{2|\mathbf{r}-\mathbf{r}_0|}{\text{L}}\right)$, where $\mathbf{r}_0$ is the center of deformed region and L is the system length of the disordered scattering region, and the varying site potential takes the similar form of $u(\mathbf{r}) = U\left(1 - \frac{2|\mathbf{r}-\mathbf{r}_0|}{\text{L}}\right)$. In our calculations, the conductance is averaged over 1000 ensembles at each point.

We first focus on the tri-layer graphene which shows the most pronounced positive piezoconductive effect. As illustrated in Fig. 3a, the tri-layer structure is constructed by mono-layer graphene in a Bernal ABA-stacking manner. Its conductance includes contributions from electron transport in both horizontal and vertical directions, and $g_{\text{max}}$ is determined by the strain-induced competition between the intralayer hopping and interlayer interactions. In the presence of an applied pressure from the AFM tip, the direct consequences on the multi-layer graphene include: 1) the extension of the lattice constant in the lattice deformed area that correspondingly decreases the intralayer nearest-neighbor hopping, and 2) the local compression of the interlayer separation in the lattice deformed area that effectively modulates the interlayer interactions, *i.e.* increasing the interlayer hopping and inducing the local site energies. By applying the non-equilibrium Green's function technique and Landauer-Büttiker formula (see details in Methods section) on a tri-layer graphene ribbon with length of 100nm and width of 13nm, we theoretically study the relation between the maximum relative conductance change $g_{max}$ and the applied strain $\varepsilon$ for tri-layer graphene. It is noteworthy to mention that all our obtained data are used to qualitatively explain and understand the underlying physical origin of the positive piezoconductive effect, and for the better presentation we have chosen certain parameters to compare with the experimental observations. Our numerical results show that $g_{max}$ is always positive and increases with increasing strain, as plotted in the blue dotted line in Fig. 3b. The experimental data are plotted in the same graph, showing the similar tendency. The non-linear feature indicates for higher strain $g_{max}$ increases slower, and suggests the relative contribution from interlayer modulation becomes less dominant. This model also explains why the piezoconductive property of bi- and multi-layer graphene is distinct from mono-layer graphene. For strained mono-layer graphene, negative peizoconductance is observed because no interlayer modulation but only intralayer contribution is present.

**Discussion**

To further understand the intriguing phenomenon that tri-layer graphene shows the most pronounced positive piezoconductive effect, we have performed first-principles calculations to study the lattice configuration of different multi-layer graphene devices under strain. Our first-principles calculations were performed using the projected-augmented-wave method as implemented in the *Vienna ab initio simulation package* (VASP)[30] with details described in the Methods section. Fig. 4a shows how the local lattice configuration is altered by the same local strain in different multi-layer graphene systems. For the bi-layer graphene, one can observe that both the top and bottom layers exhibit the structural deformations with different amplitudes. We further plot the strain-induced lattice variation $\Delta d$ between the nearest two layers as a function of the layer number *n* in the Fig. 4b. Along with the increase of the layer number, *e.g.* from bi-layer to up to hexa-layer, only the top three layers show noticeable structural deformations. Therefore the positive piezoconductive effects for multi-layer graphene systems due to the interlayer modulations should be dominated by the top three layers.

We can now introduce a characteristic factor, the piezoconductive factor $\gamma$, to describe the piezoconductive properties of bi- and multi-layer graphene devices:

$$g_{\max} = \gamma \cdot \varepsilon^{1/2} \quad (6)$$

The piezoconductive factor $\gamma$ has the same sign as piezoconductive effect. The square root dependence on strain is phenomenological but describes the experiment and theoretical data well and allows us to parameterize each $g_{\max}(\varepsilon)$ curve by a single value $\gamma$. By applying the same non-equilibrium Green's function technique as used in studying the tri-layer graphene, we calculate the $g_{\max} - \varepsilon$ relation for graphene with various layer number $n$ ($n$ = 2 to 6) (see detailed parameters in Supplementary Table 1), and extract the values of $\gamma$ by using the above definition. The results are shown by the blue dashed line in Fig. 4c. The values of $\gamma$ are positive, with the maximum value around 1.6 for tri-layer graphene. To compare the results with experimental observation, we plot the experimental data in the same graph (red dotted line), and find they are consistent with theoretic results. For bi- and tri-layer graphene, $\gamma$ increases with $n$, suggesting the enhancement of the contribution from strain-induced interlayer modulation. But while $n$ increases further, $\gamma$ decreases, suggesting the interlayer modulation is dominated by the top three layers and additional parallel conduction paths from the extra layers suppress the positive piezoconductive effect. Here we note that, the roles of external effects (see Supplementary Figs 2-5 and Supplementary Note 2 for details), the stacking order (Supplementary Figs 6-8 and Supplementary Note 3) and back gate (Supplementary Fig. 9 and Supplementary Note 4), as well as finite size effect in the theoretical calculations (Supplementary Figs 10-11 and Supplementary Note 5), have been carefully explored in our work. Although the observation of negative piezoconductive effect in multi-layer graphene was reported previously[31], it is likely related to the devices' unusually large contact resistance, rather than the intrinsic piezoconductive properties of multi-layer graphene as revealed in this manuscript. In summary, we have systematically studied the piezoconductive effect of suspended graphene with different number of layers. In contrast to the negative piezoconductance observed in mono-layer graphene, positive piezoconductance is observed in bi- and multi-layer graphene, with tri-layer graphene showing the most pronounced response. It can be explained by the model of strain-induced competition between electronic interlayer coupling and intralayer transport. This model is further confirmed by numerical calculations based on non-equilibrium Green's function method. Our results enrich the understanding of electromechanical properties of graphene and underscore their potential applications on the field of NEMS and flexible electronics.

## Methods

### Suspended graphene device preparation and characterization

The suspended graphene membranes are obtained by using mechanical exfoliation method on the pre-defined trenches on 300nm-thick $SiO_2$ wafers. The trenches are defined by standard photolithography method, followed by dry etching in an ICP system, where $CH_4$ and $CHF_3$ are used as etching gases. The typical width of the trenches is 3μm and the depth is around 250nm. The number of layers of graphene membranes is first identified by an optical microscopy and further confirmed by Raman spectroscopy. To avoid common wet process induced device performance degradation and yield loss, the electronic devices are fabricated by using a home-made shadow mask method[24]. The electrodes are made of 5nm Ti covered by 50nm Au.

### Setup of Pressure-modulated Conductance Microscopy (PCM)

A Bruker Multimode 8 Atomic Force Microscopy (AFM) is used to build up the PCM setup. A non-conducting AFM tip is used to apply adjustable local pressure on the top of the suspended graphene, yielding topography/height image of the strained graphene. The conductance/resistance of the device is measured via lock-in technique and monitored simultaneously as an external input of the AFM. The comparison of the conductance/resistance image and topography image offers detailed piezoconductive information of the suspended graphene membranes. The scanning procedure is done in contact mode with a Bruker NP-S10 tip, of which the Young's modulus is 0.24Nm$^{-1}$. The detailed experimental setup is schematically shown in Fig. 1b.

### Details of non-equilibrium Green's function simulation

Using multi-probe Landauer-Büttiker formula, the current in the lead $p$ (either real or virtual lead) can be expressed as:

$$J_p = (2e^2/h) \sum_{q \neq p} T_{pq}(V_p - V_q) \qquad (7)$$

where $V_{p/q}$ is the spin-independent bias in the lead $p/q$. The electronic transmission coefficient from lead $q$ to lead $p$ is calculated as $T_{pq} = \text{Tr}[\Gamma_p G^r \Gamma_q G^a]$, in which the line-width function $\Gamma_p$ is defined as $\Gamma_p = i[\Sigma_p^r - \Sigma_p^a]$, and the retarded and advanced Green's function are given by $G^r = [G^a]^\dagger = [(E_F + i\eta)I - H - \sum_p \Sigma_p^r]$, where $I$ is the unit matrix with the same dimension as that of $H$. The retarded and advanced self-energy due to the coupling to all the real leads can be obtained numerically[32]. For the virtual leads, we assume $\Sigma_p^r = -i\Gamma_d/2$ and the dephasing strength $\Gamma_d$ is fixed to $0.01eV$. In our simulations, a small external bias is applied between the left and right lead with $V_L = -V_R = 0.5V$. For dephasing effect, electrons lose the phase memory by entering and leaving the virtual leads. Thus, for each virtual lead $i$, the current has the constraint that $J_i = 0$, which ensures the current conservation. Combining the above equation of $J_p$ together with all boundary conditions for the real and virtual leads, the voltage $V_p$ and current $J_p$ in each real lead can be obtained.

**Details of First-principles calculation**

The generalized gradient approximation (GGA) combined the Vander Waals correction with the DFT-D2 method of Grimme is used. The kinetic energy cutoff is set to be 500eV. During the structure relaxation, the edge atoms and the probed atoms are not allowed to relax while the others are. All parameters are chosen to converge the forces to be less than $0.01eV\text{Å}^{-1}$. The first Brillouin-zone integration is carried out by using the 3×3×1 Gamma-centered grids. A vacuum buffer space of 20Å is set to prevent the interaction between adjacent slabs.

## Acknowledgements

This work was supported in part by the National Key Basic Research Program of China (2015CB921600, 2011CB922103, 2013CBA01603), the National Natural Science Foundation of China (11374142), the Natural Science Foundation of Jiangsu Province (BK20130544, BK20140017), the Specialized Research Fund for the Doctoral Program of Higher Education (20130091120040), Fundamental Research Funds for the Central Universities, and the Collaborative Innovation Center of Advanced Microstructures. K. W., Y. R. and Z. Q. are supported by 100 Talents Program of Chinese Academy of Sciences, NNSFC (11474265) and Anhui Provincial Natural Science Foundation, The Supercomputing Center of USTC is gratefully acknowledged for the high-performance computing assistance.

## Author contributions

F. M. conceived the project and designed the experiments. K. X., W. B., E. L., M. W., Y. F. and J. Z. carried out device fabrication and electrical measurements. K. W., W. Z., Y. R., Z. Q. carried out numerical simulations. Z. L. and W. Z. carried out Raman spectroscopy measurements and analysis. F. M., M. S. F., B. W., Z. Q., K. X., K. W. and W. Z. carried out data analysis and interpretation. F. M., M. S. F., Z. Q., K. X. and K. W. co-wrote the paper with all authors contributing to the discussion and preparation of the manuscript.

## Competing financial interests

The authors declare that they have no competing financial interests.

## Figure Captions

**Figure 1. Suspended graphene device and PCM setup. a**, Optical microscope image of a four-terminal suspended bi-layer graphene device, which was fabricated by home-made shadow mask method. Insert: SEM image of a suspended device. The scale bar is 3μm. **b**, Schematic setup of Pressure-modulated Conductance Microscopy (PCM) which performs piezoconductive measurements on suspended graphene devices.

**Figure 2. Layer number dependent positive piezoconductive effect of graphene. a-d**, Left panels are the line traces of the topography (tip position) image of graphene devices from mono-layer (a) to tetra-layer (d) with applied strain $\varepsilon = 0.54‰, 0.51‰, 0.33‰, 0.20‰$ respectively. Right panels are the line traces of the corresponding relative conductance change $g$ (to the undisturbed conductance with no local pressure applied). Mono-layer device shows negative piezoconductive effect (conductance drops upon local pressure applied). Bi-, Tri- and Tetra-layer devices shows positive piezoconductive effect (conductance jumps upon local pressure applied) with the most pronounced effect in Tri-layer device. **e**, The plot of the maximum relative conductance change (when AFM tip approaches the center of the suspended membranes) $g_{\max}$ as a function of strain $\varepsilon$ for various suspended graphene devices (layer number $n$=1, 2, 3, 4, 6 represented by different colors).

**Figure 3. Theoretical model and numerical calculations of tri-layer graphene. a**, Schematic of the lattice structure change of Bernal (ABA) stacking tri-layer structure due to a vertical load **F** applied. **b**, The maximum relative conductance change $g_{\max}$ as a function of strain $\varepsilon$ for tri-layer graphene. The red squares are the experimental data while the blue cycles are the numerical simulation result.

**Figure 4. Simulation results of the dependence on layer number. a,** The structurally relaxed configuration for different multi-layer graphene (layer number $n = 2 - 6$ from left to right) in the presence of the same strain strength. **b**, The strain-induced lattice variation $\Delta d$ between the nearest two layers as a function of the layer number $n$. **c**, The dependence of the piezoconductive factor $\gamma$ on layer number $n$, with experimental data represented by red squares and simulation results represented by blue squares. The error bars of experimental data originate from the fitting process, while the error bars of simulating results originate from different disorder configurations in the numerical calculations.

Figure-1 (Xu *et. al.*)

a 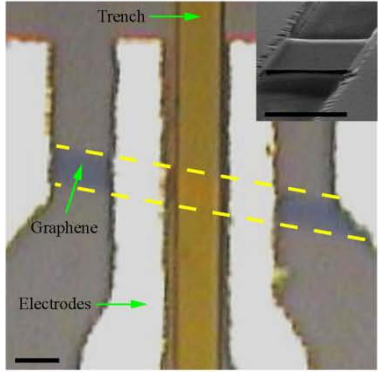 b 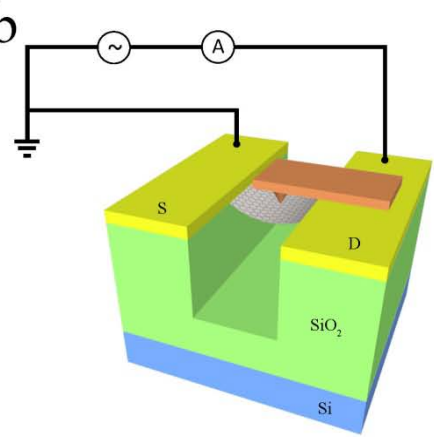

# Figure-2 (Xu *et. al.*)

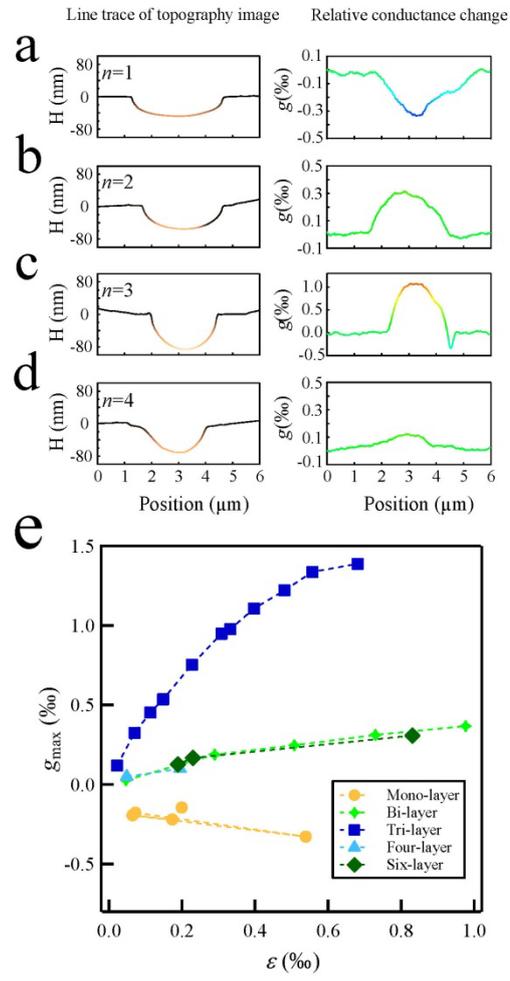

Figure-3 (Xu *et. al.*)

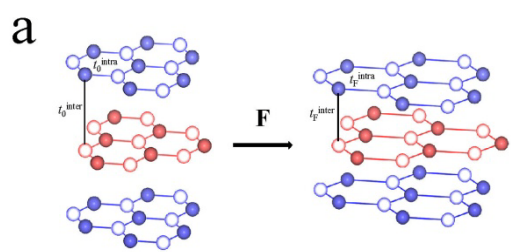

a

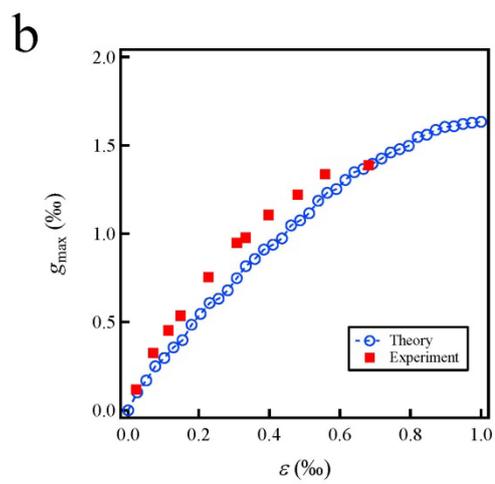

b

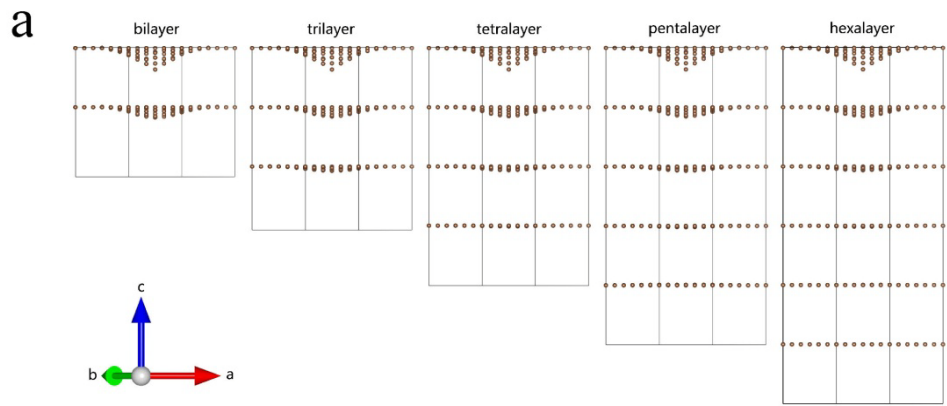

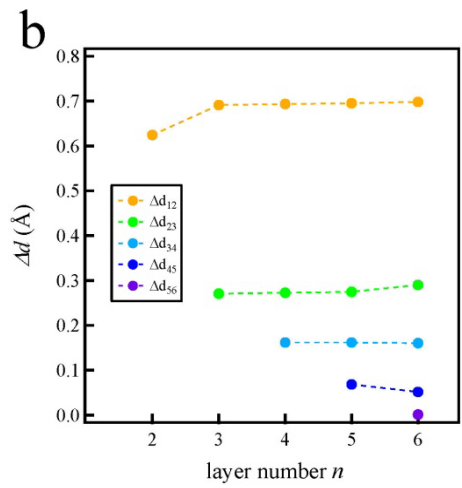
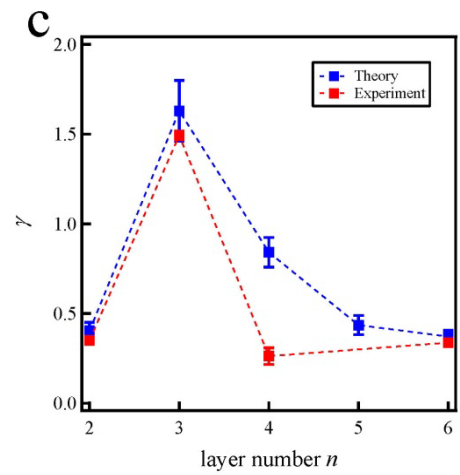

**Supplementary Figures**

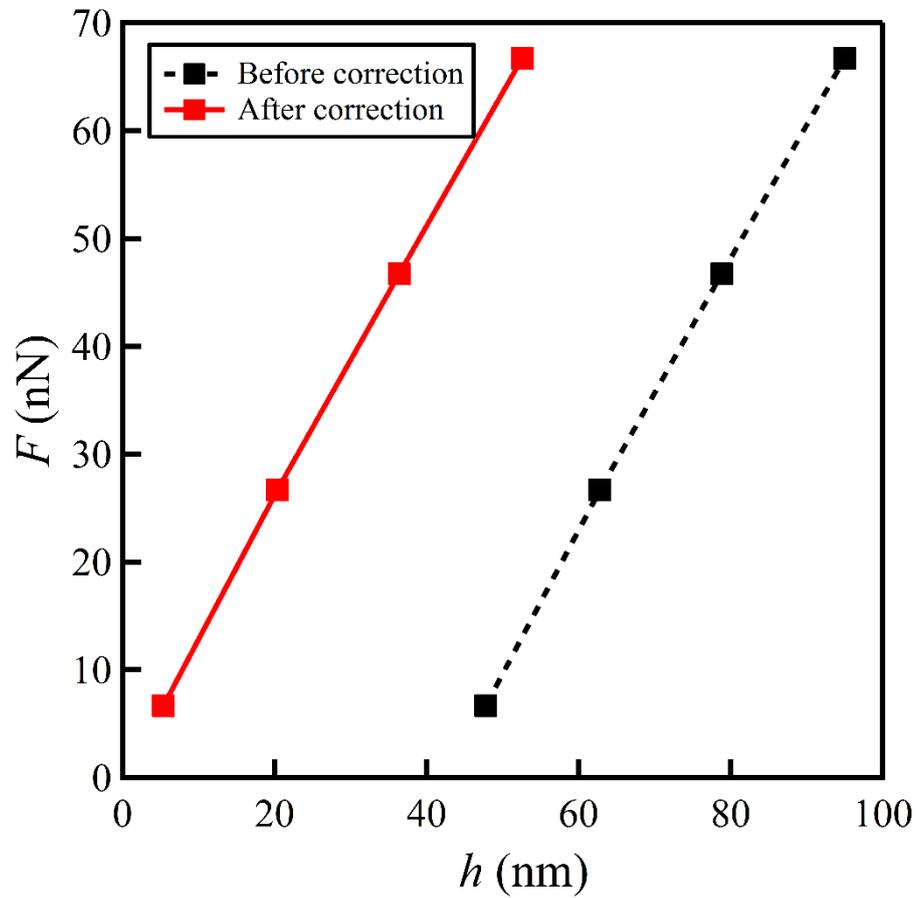

**Supplementary Figure 1: Load versus elastic deformation of a typical device.** The black squares represent the original data points, and the red squares represent the data points after subtracting the intercept.

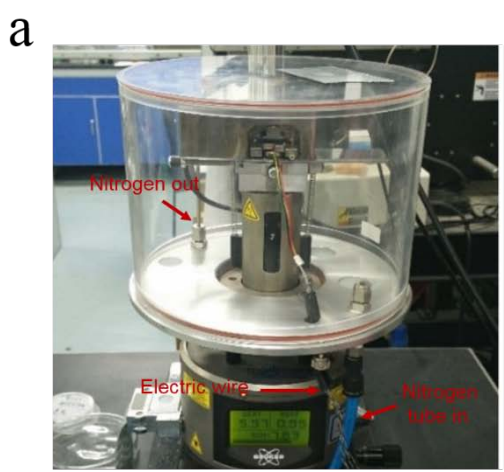 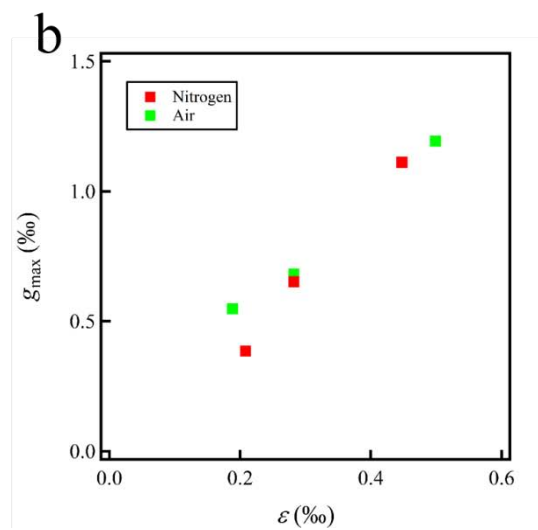

**Supplementary Figure 2: Piezoconductive effect experimented in nitrogen and air.** The maximum relative conductance change $g_{max}$ as a function of strain $\varepsilon$ for trilayer graphene. The experimental data acquired in nitrogen and air were shown by red and green squares respectively.

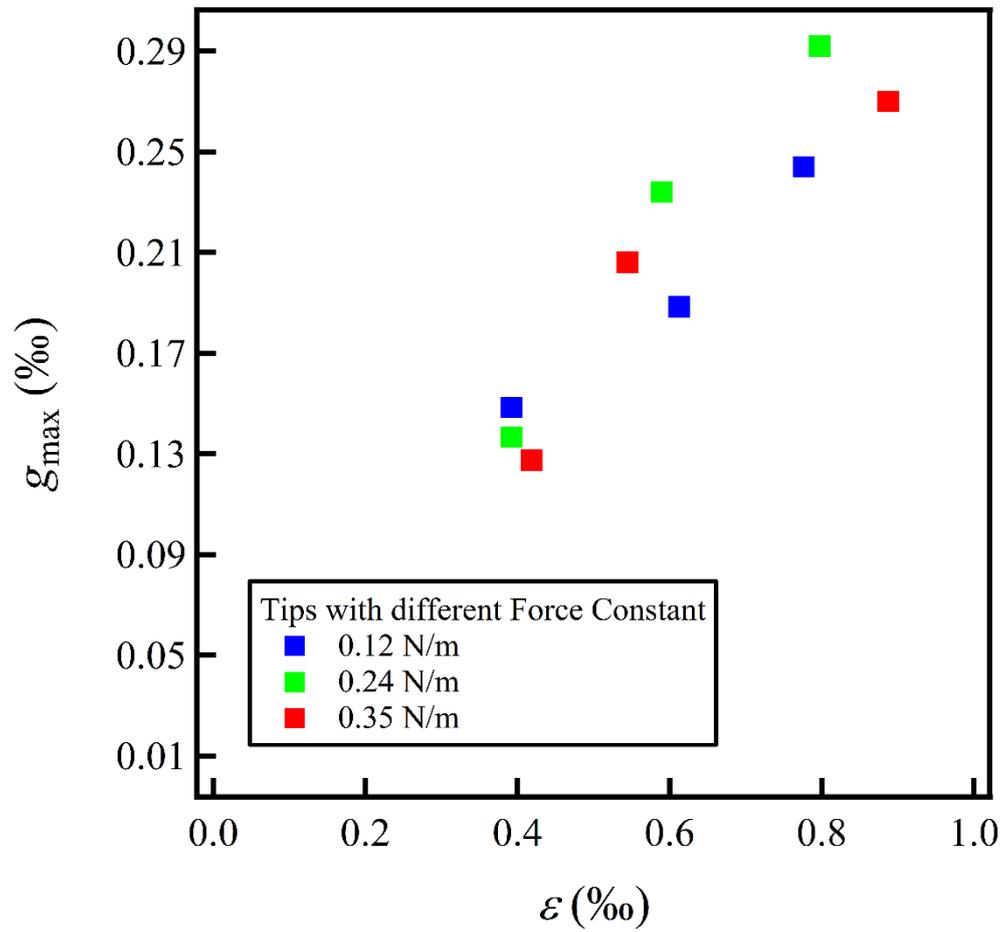

**Supplementary Figure 3: Positive piezoconductive effect for three different kinds of AFM tips.** The maximum relative conductance change $g_{max}$ as a function of strain $\varepsilon$ for a bilayer graphene. The experimental data for three different kinds of tips were shown by different color markers.

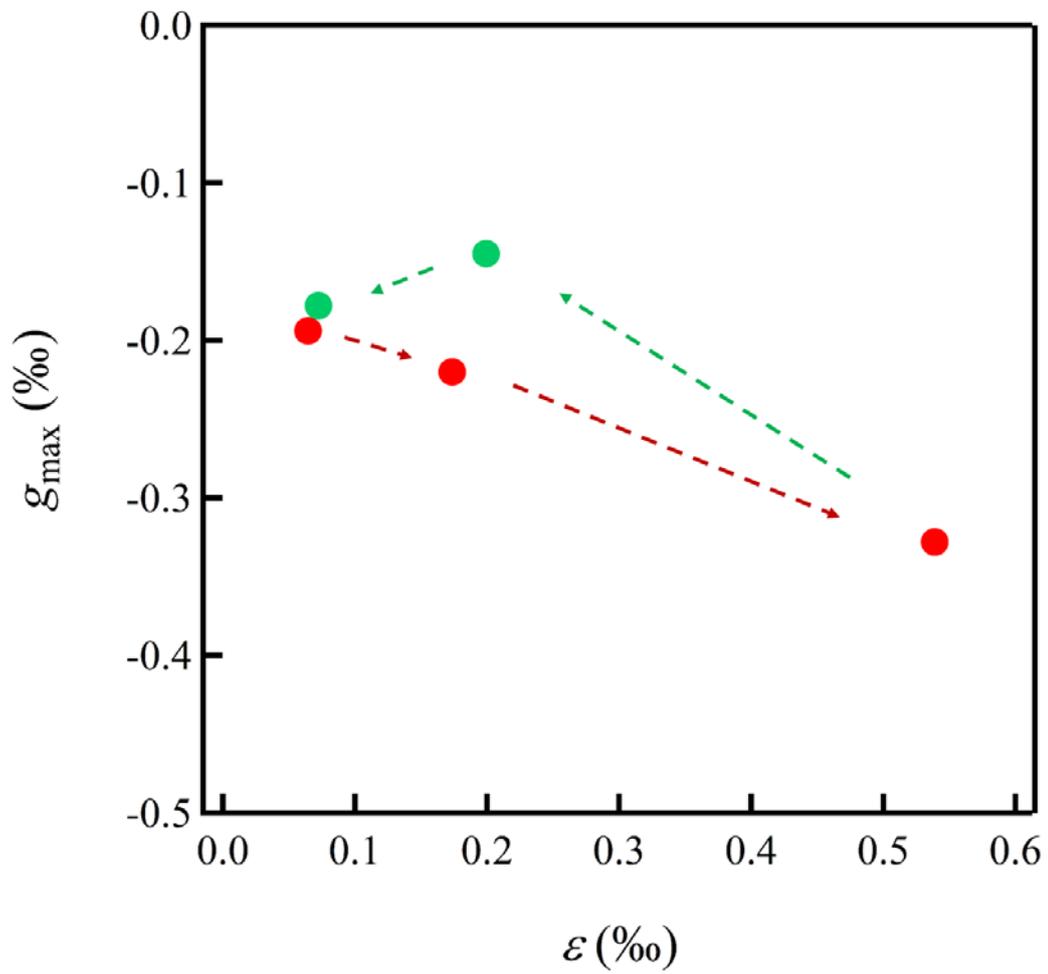

**Supplementary Figure 4: Replot of $g_{max}$ as a function of strain $\varepsilon$ for trilayer graphene data in Figure 2e displayed in the main text.** The measurement order is labelled near the data points and green dots represent the data from the reverse measurement.

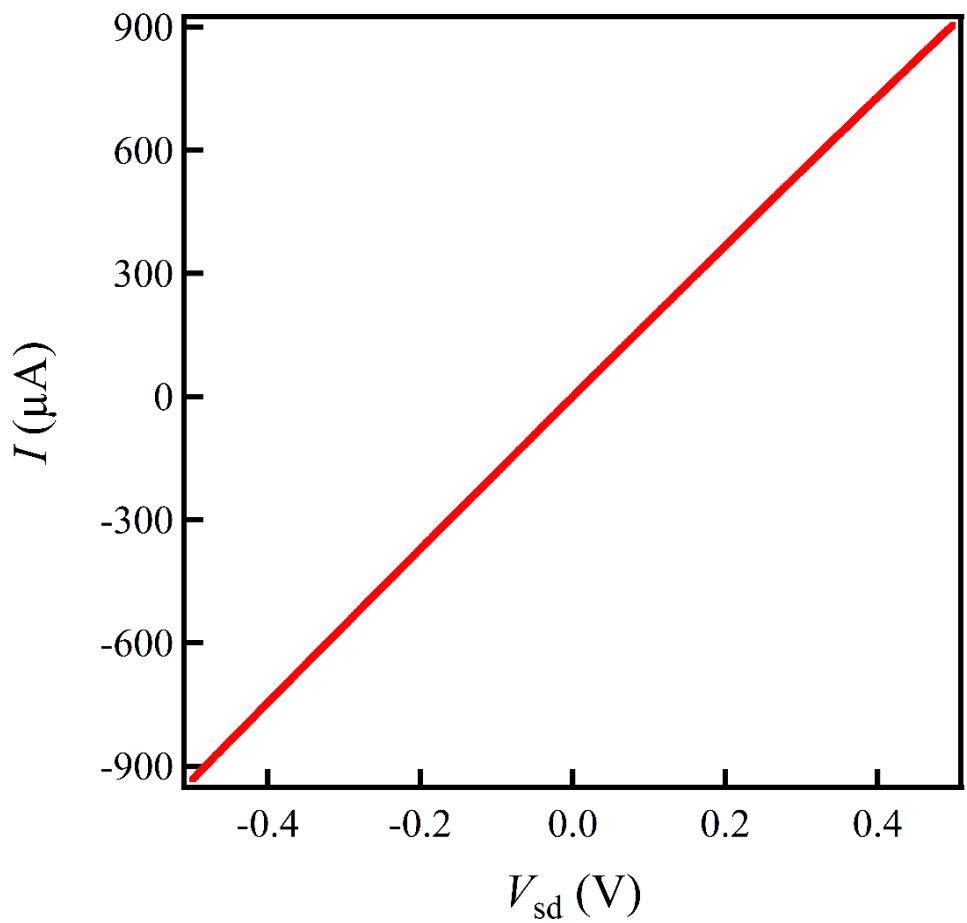

**Supplementary Figure 5: *I-V* curve of the trilayer graphene device.** $V_{sd}$ is varied from -0.5 V to 0.5V.

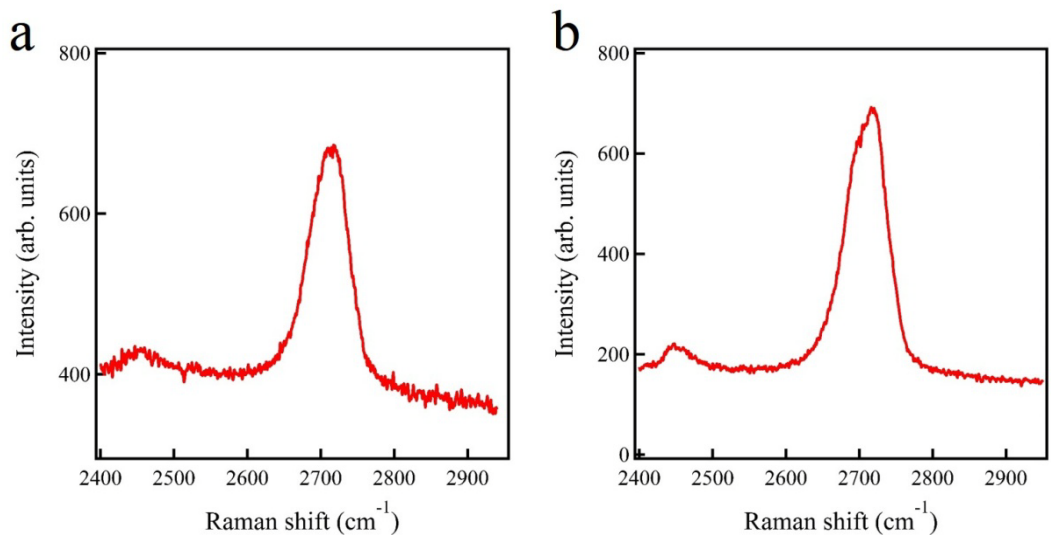

**Supplementary Figure 6: Raman spectra of the tetralayer (a) and hexalayer graphene (b).**

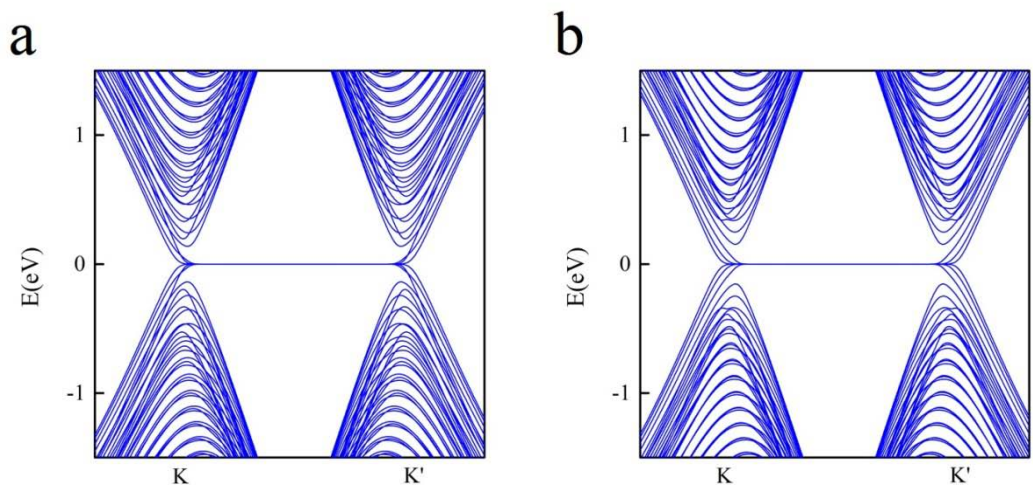

**Supplementary Figure 7: Band structure of trilayer graphene nanoribbons with different stacking order.** (a) ABA stacked trilayer graphene nanoribbons; (b) ABC stacked trilayer graphene nanoribbons.

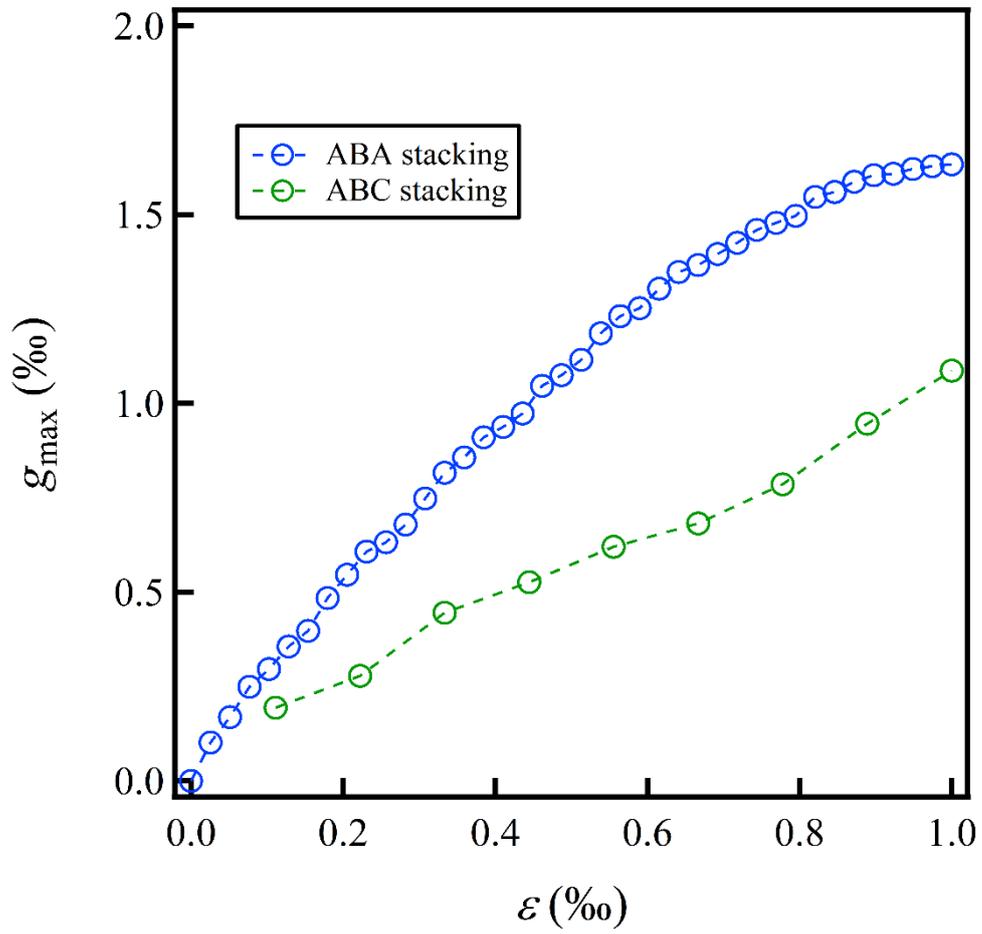

**Supplementary Figure 8: The maximum relative conductance change $g_{max}$ as a function of strain $\varepsilon$ for ABA and ABC stacked trilayer graphene.**

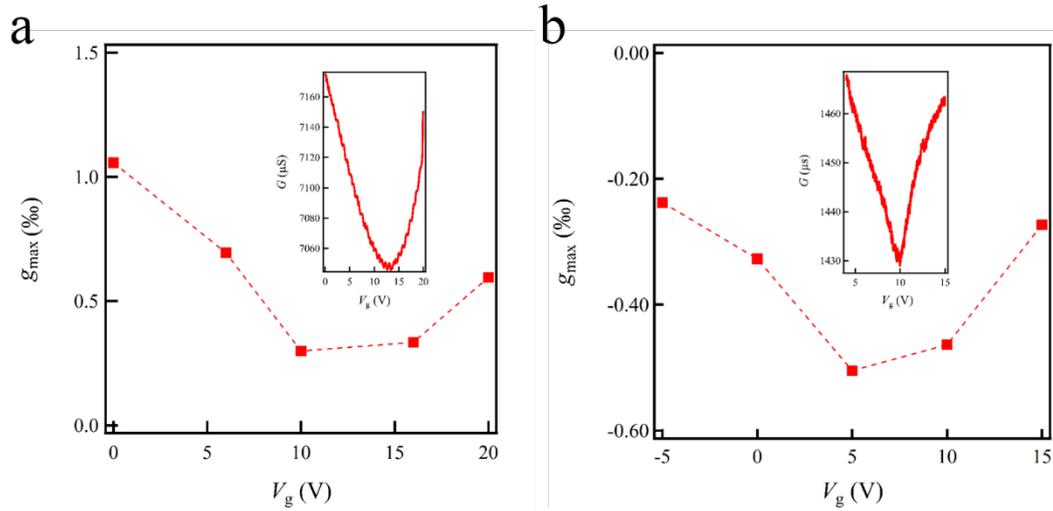

**Supplementary Figure 9: Dependence of positive piezoconductive effect on back gate voltage of trilayer (a) and monolayer (b) graphene device.** $G$ versus $V_g$ data are inserted.

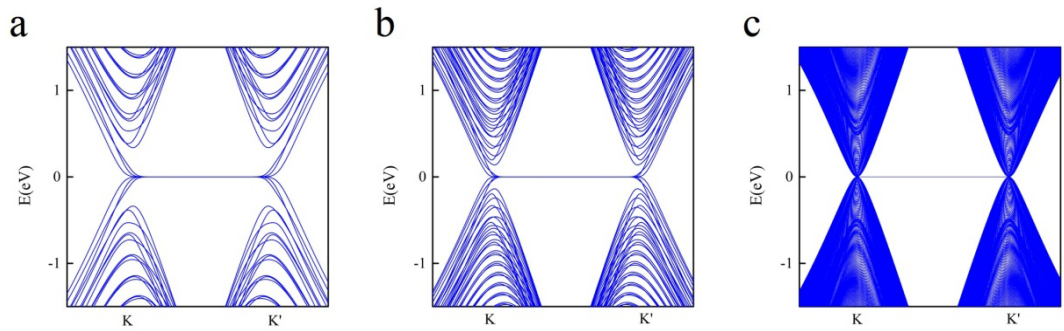

**Supplementary Figure 10:** Comparisons of the band structures of the zigzag trilayer (ABA stacking) graphene nanoribbons with different widths, i.e., 6 nm (a), 13 nm (b), and 100 nm (c).

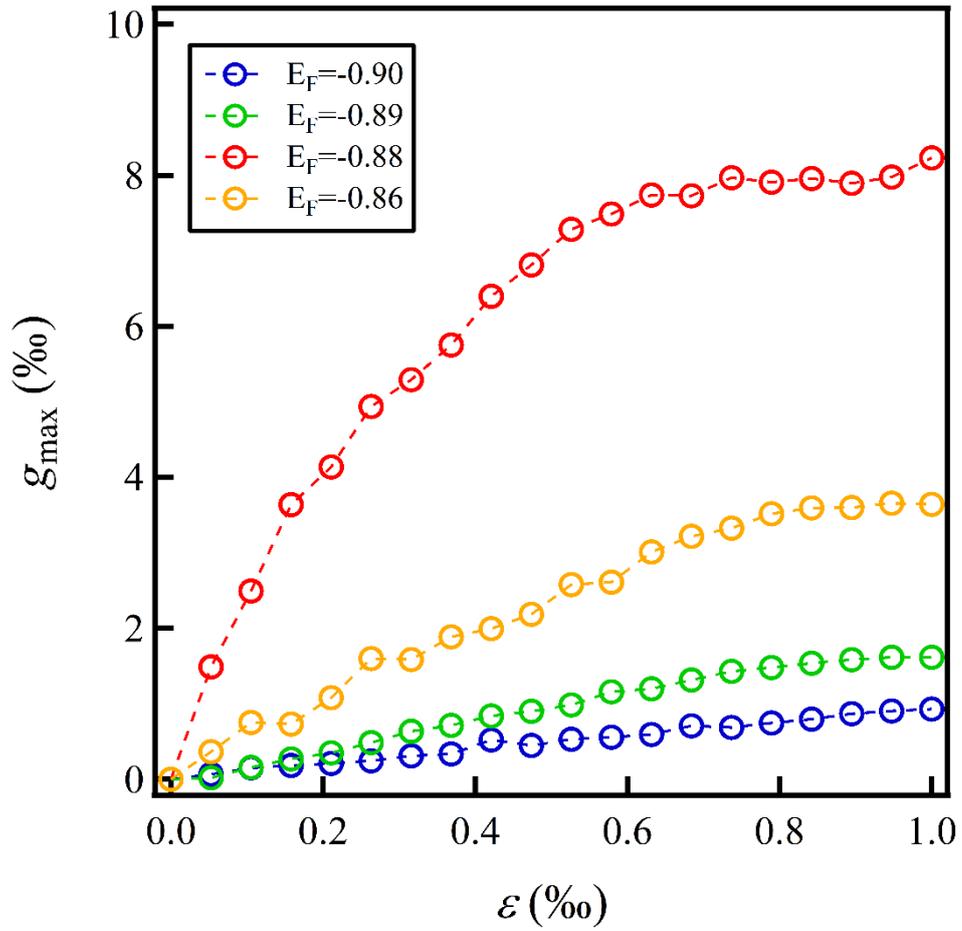

**Supplementary Figure 11: Piezoconductive effects *vs* different Fermi levels for trilayer graphene**. Scattering area is 13 *nm* × 100 *nm*.

## Supplementary Notes

**Supplementary Note 1: The suspended bridge model in small strain regime**

As described in the suspended bridge model [1], the dependence of the load $F$ on the maximum strain-induced deflection $h$ can be expressed by:

$$F = \frac{8wE^{2D}}{l^3}h^3 + \frac{8w\sigma^{2D}}{l}h \quad (1)$$

where $w$ is the width, $E^{2D}$ is the two-dimensional Young's modulus, $\sigma^{2D}$ is the residual tension, $h$ is the maximum strain-induced deflection, and $l$ is the length of the suspended graphene. In the small strain regime, the load $F$ is dominated by the second term. Experimentally, $F$ can be calculated based on the deflection voltage value set in the contact mode of AFM. A typical measured $F$-$h$ curve is shown in Supplementary Figure 1 (black dots). The clear linear dependence indicates the validation of the suspended bridge model in our experimental setup. The slope is proportional to the residual tension $\sigma^{2D}$, which is found to be quite small in our devices, with typical values ranging between 0.02N/m and 0.14N/m.

The intercept at the *x*-axis equals to the height of the tip at which the force begins to rise. For the devices studied, the intercept varies between 6 nm and 42 nm, which may be attributed to the randomness of the mechanical exfoliation process. To get the precise strength of the strain, the actual maximum strain-induced deflection $h$ needs to be corrected by subtracting this height. As shown in Supplementary Figure 1, the intercept is determined to be about 42nm, and the corrected $F$-$h$ curve is denoted by the red dots.

**Supplementary Note 2: Role of external effects**

Considering our experimental setup, it is important to explore the role of several external effects that could affect the experimental observations. The details are discussed in the following:

**Influence of ambient air**

Our measurements were performed in ambient air, which may induce device stability issue in certain circumstances. We have investigated the influence of ambient air on the measurements very carefully. We first added an atmospheric hood (an accessory of the Bruker Multimode 8 AFM we used) to the setup of the pressure-modulated conductance microscopy (PCM), with a photograph shown in Supplementary Figure 2a. The piezoconductive measurements were then performed on a same tri-layer device in both the ambient air and pure nitrogen environment. The positive piezoconductive effect was observed in both cases, with the obtained data shown in Supplementary Figure 2b. It clearly shows that the data acquired from the pure nitrogen environment (red square symbols) are nearly collapsed with those acquired in the ambient air (green square symbols). This strongly indicates that the random doping from the external environment has negligible influence on our results.

**AFM tip-induced parasitic capacitor effect**

In our consideration, the graphene membranes are in contact with a non-conductive AFM tip which could induce parasitic capacitor effect. If the tip is randomly charged (*i.e.*, full of random locally charged sites), it would form a capacitor with the graphene sheets, which would affect the carrier density and result in a graphene resistivity change. In order to explore the influence of such effect, we have repeated the piezoconductive measurements on a same bi-layer device, but using three different types of AFM tips. These tips have various force constants (0.12N/m, 0.24N/m, and 0.35N/m respectively), distinct tip geometries and therefore random charge distributions. The experimental results are displayed in Supplementary Figure 3. The observed positive piezoconductive effect follows a same curve, indicating that the tip-induced parasitic capacitor effect is also negligible in our experiments.

**Local tearing**

If there is a local tearing, the resistance of the suspended graphene would increase, which could affect the piezoconductive measurements (especially in the case of single layer graphene). To exclude that local tearing plays a role in our measurements, the data of the monolayer graphene in Figure 2e of the main text were taken in a measurement of hysteresis. In Supplementary Figure 4, we have clearly marked the order of the data points taken in the measurement, where two data points (green symbols) were obtained during the reverse measurement. Since the resistance change is reversible, it is reasonable to conclude that the local tearing does not occur during our measurement. Furthermore, during our measurements, for each line scan, we also monitored the conductance chance of the device in real time before and after applying the stress. If there is a local tearing due to the deformation from the AFM tip, the resulting conductance should not be able to return to its original value (due to a resistance increase) when the AFM tip re-approaches the electrodes. As shown in Figure 2a of the main text, one can find that the conductance line trace of the monolayer graphene is rather symmetric, which indicates that there is no local tearing during our measurements.

**Self-heating effect**

The self-heating effect may also influent the conductivity of graphene. In our experiment, we have applied a rather small source-drain bias (<0.1V) on the graphene devices by using a Lock-In Amplifier. The self-heating effects should be negligible in our measurement. To rule out this effect, we have tuned the source-drain voltage difference from -0.5V to 0.5V to measure the current of a typical tri-layer graphene device. As displayed in Supplementary Figure 5, the I-V curve shows a linear characteristic, providing a strong evidence to exclude the self-heating effect.

**Supplementary Note 3: Role of the stacking order**

The Raman spectra of the four- and six-layer graphene devices are displayed in Supplementary Figure 6a and 6b respectively. After comparing with the literatures [2], we can conclude that the four-layer graphene is stacked in an ABAB order. However, it is unclear of the stacking order of the six-layer graphene device based on the obtained Raman spectra, due to a lack of related studies in literatures.

In order to explore whether the stacking order plays a crucial role in the piezoconductive effect of the multi-layer graphene, we plot the band structures of both the ABA- (a) and ABC-stacked (b) tri-layer graphene nanoribbons in Supplementary Figure 7, which are totally different [3]. Nevertheless, based on the analysis of the physical origin of the positive piezoconductive effect, we believe that the ABC-stacked tri-layer graphene should have the similar positive piezoconductive effect because of the strain-induced alterable interlayer interaction. In Supplementary Figure 8, we have theoretically verified that the ABC-stacked tri-layer graphene can also exhibit a similar positive piezoconductive effect.

**Supplementary Note 4: Back-gated suspended graphene**

Back-gated suspended graphene device is a rather complicated system when experiencing the combined influences from the back gate, the strain and the geometric deflection. These parameters influent the conductivity of the graphene flake jointly since they are correlated with each other. In an ideal case, tuning the back gate only shifts the Fermi level of the whole system. However, in the real case, applying a back gate voltage on a suspended graphene device induces inhomogeneous carrier distribution and many parasitic effects due to the non-negligible geometric change. One example is the parasitic capacitive gating effect, i.e. the carrier density in the graphene increases as the graphene membranes are pushed closer to the back gate oxide during the experiments. Another example is the attractive electrostatic force-induced deflection effect [4] which alters the strain, the geometric deflection, and thus the parasitic capacitive gating effect as well. Thus, in such a gated suspended graphene system, it is highly challenging to distinguish the role of the varying Fermi level from that of these parasitic effects.

Experimentally, we have repeated the PCM experiments by applying different back gate voltages on both trilayer (see Supplementary Figure 9a) and monolayer graphene devices (see Supplementary Figure 9b). As displayed in the inset of Supplementary Figure 9a, the charge neutrality point of the tri-layer graphene is about 13V. We then measured the piezoconductive effect of this device under different back gate voltages (*i.e.*, $Vg$ =0, 6, 10, 16, and 20V), and obtain the corresponding $g_{max}$. One can observe that the positive piezoconductive effect becomes more pronounced when $Vg$ is away from the charge neutrality point, and the relation of $g_{max}$-$V_g$ follows a parabolic character. For the monolayer graphene device, the experimental results show a similar parabola-shaped dependence as displayed in Supplementary Figure 9b. Such results suggest that for the higher gate voltages (negative and positive), the parasitic capacitive gating effect enhanced the conductivity, resulting in more pronounced positive piezoconductive effect in tri-layer graphene, or weakened negative piezoconductive effect in monolayer graphene.

We have also theoretically investigated the piezoconductive effect as a function of the Fermi level, and found that the finite-size effect plays a significant role in determining the relationship between the piezoconductive effect and the Fermi level. There is a detailed discussion in the following note "**Supplementary Note 5: Finite size effect in the theoretical calculations**".

**Supplementary Note 5: Finite size effect in the theoretical calculations**

In our numerical calculations, we have considered a system size of 13 *nm* × 100 *nm* (about $3 \times 10^5$ atoms for trilayer graphene), and our numerical results are able to qualitatively explain the experimental findings. However, due to the limitation of the computational capacity, it is impossible to model an extremely large system that is comparable to the practical devices. Therefore, it is unavoidable to introduce the finite size effects in the numerical calculations. In Supplementary Figure 10, we plot the band structures of the zigzag trilayer graphene nanoribbons with different ribbon widths, *i.e.*, 6 *nm* (a), 13 *nm* (b) and 100 *nm* (c). One can see that there is a completely distinct energy spectra around the charge neutrality point (*i.e.*, $E_F=0$ in our calculation). The weakness of the finite size effect can be solved by setting the Fermi level to be slightly away from the charge neutrality point $E_F=0$ in our numerical calculations, and the difference of the energy spectra near the charge neutrality point does not qualitatively affect our results. It is noteworthy that all the samples in our experiments are p-doped, leading to the fact that the Fermi level is naturally away from the charge neutrality point.

In addition, the number of transverse modes for a given Fermi level of the small systems (*e.g.* 6 *nm* and 13 *nm*) is significantly different from the large systems (*e.g.* 100 *nm*). As displayed in Supplementary Figure 11, the maximum relative conductance change $g_{max}$ is highly dependent on the Fermi level due to the density of transverse modes is closely rely on the Fermi level in the numerical calculations. However, one can clearly observe that the piezoconductive effects for different Fermi levels are all positive in the tri-layer graphene. This shows that our numerical calculation can be able to qualitatively capture the physical origin of the positive piezoconductive effect. In our numerical calculations, we have better explained the experimental findings by choosing suitable parameters (*e.g.* the Fermi level). However, we want to stress that our theory just aims to qualitatively reveal the physical origin of the piezoconductive effect, but not to exactly fit the experimental data.

## Supplementary Tables

**Supplementary Table 1: Fitting parameters for different layer number.** $U = \eta\varepsilon$ and $\delta t = \chi\varepsilon$ in units of eV, with the fitting parameters $\eta$ and $\chi$ for the bi-layer, tri-layer and the top three layers when the layer number $n > 3$ (both terms of other layers are assumed to be zero for simplicity).

| $n$ | 2 | 3 | 4 | 5 | 6 |
|---|---|---|---|---|---|
| $\eta$ | 20 | 15 | 15 | 15 | 15 |
| $\chi$ | 200 | 50 | 50 | 50 | 50 |

## Supplementary References